\documentclass[cameraready]{Interspeech}

\title{Evidence Subspace Projection: Measuring How Much Evidence Explains Deepfake Detection in Self-Supervised Speech Models}

\author[affiliation={1}]{Yixuan}{Xiao}
\author[affiliation={1}]{Cheng-Wei}{Lin}
\author[affiliation={2}]{Xin}{Wang}
\author[affiliation={3}]{Yassine}{El Kheir}
\author[affiliation={3}]{Arnab}{Das}
\author[affiliation={3}]{Tim}{Polzehl}
\author[affiliation={4}]{Sebastian}{M\"oller}
\author[affiliation={1}]{Ngoc Thang}{Vu}

\address{
    $^1$ University of Stuttgart, Germany \\
    $^2$ National Institute of Informatics, Japan \\
    $^3$ German Research Center for Artificial Intelligence (DFKI), Germany\\
    $^4$ Technical University of Berlin, Germany
}

\email{yixuan.xiao@ims.uni-stuttgart.de}


\usepackage{comment}
\usepackage{amssymb}
\usepackage{amsmath}
\usepackage{colortbl}
\usepackage{booktabs}
\usepackage{multirow}
\usepackage{graphicx}

\begin{document}

\maketitle

\begin{abstract}

Self-supervised learning (SSL) models are widely used as feature extractors for state-of-the-art audio deepfake detection, 
but it remains unclear how to directly and quantitatively connect what SSL models capture to detection decisions. 
To address this gap, we propose Evidence Subspace Projection, a method that represents both evidence factors (e.g., attack category, codec, gender, transmission) and authenticity labels in a shared space constructed from SSL models’ neuron activation patterns. By projecting the decision vector onto each evidence subspace, we obtain a scalar ratio that quantifies the explanatory power of each evidence type.
We evaluate SSL models in raw, fine-tuned, and post-trained settings on multiple datasets. The results confirm findings from established studies, validating the proposed method, and reveal new insights into model behavior.

\end{abstract}

\section{Introduction}

Audio deepfake detection built on self-supervised learning (SSL) front-ends achieves strong in-domain performance, even with lightweight back-ends\cite{babu22_xlsr2b, pascu24_generalisable}. But this performance often fails to generalize to out-of-domain data, suggesting reliance on dataset-specific cues rather than genuine spoofing artifacts~\cite{combei25_AI4T}.
Hence understanding what information SSL representations actually rely on for detection is important both for interpretability and for improving generalization.

Prior works mostly analyze the detector as a whole~\cite{ge2022_explaining, liu24m_neuralspoofing, muller21_speechissilver, shih2024_artifacts},
conflating the SSL front-end's contribution with effects introduced by back-end design, loss function, and fine-tuning dynamics. 
Yet a frozen SSL front-end alone can achieve strong performance~\cite{pascu24_generalisable},
and training-free methods such as k-NN on SSL features suffice for deepfake source tracing~\cite{stan25_interspeech},
indicating that the front-end itself plays a key role. This motivates us to study the front-end in isolation.

We formulate two \textbf{research questions}: \textbf{RQ1}: What does a frozen front-end already capture about deepfake detection? Specifically, to what extent does its detection decision overlap with specific evidence factors such as silence structure, spectral properties, or attacker identity? \textbf{RQ2}: How do fine-tuning and post-training reshape the relationship between the detection decision and these evidence factors?

To address these, we propose \emph{Evidence Subspace Projection}. Building on the key-value memory view of Transformer FFN layers~\cite{geva2021transformer}
and property neuron analysis in speech SSL models~\cite{lin2024property}, we represent each label (e.g., bonafide, TTS) as a one-vs-rest contrast vector derived from neuron activation statistics. 
Both the detection decision and evidence factors are represented in the same neuron-activation space, enabling direct comparison.
We then project the decision axis onto each evidence subspace and measure the fraction of explained variance, providing a quantitative, per-factor measure of explanatory power.

Our \textbf{contributions}\footnote{\url{https://github.com/XIAOYixuan/ESP}} are: 
(1) the first neuron-level analysis applied to audio deepfake detection, studying SSL front-ends in isolation;
(2) Evidence Subspace Projection, a method that quantifies how much each evidence factor explains the detection decision;
(3) analysis across two SSL models, three training conditions, six datasets, and nine factors, revealing that: frozen models already encode dataset-specific shortcuts; within-spoof alignment diagnoses the model's vulnerability; training data homogeneity causes signal-level confounds while diversity de-correlates them; and post-training suppresses most dependencies, though silence structure remains persistent.

\begin{figure}[t]  
    \centering
    \includegraphics[width=\columnwidth]{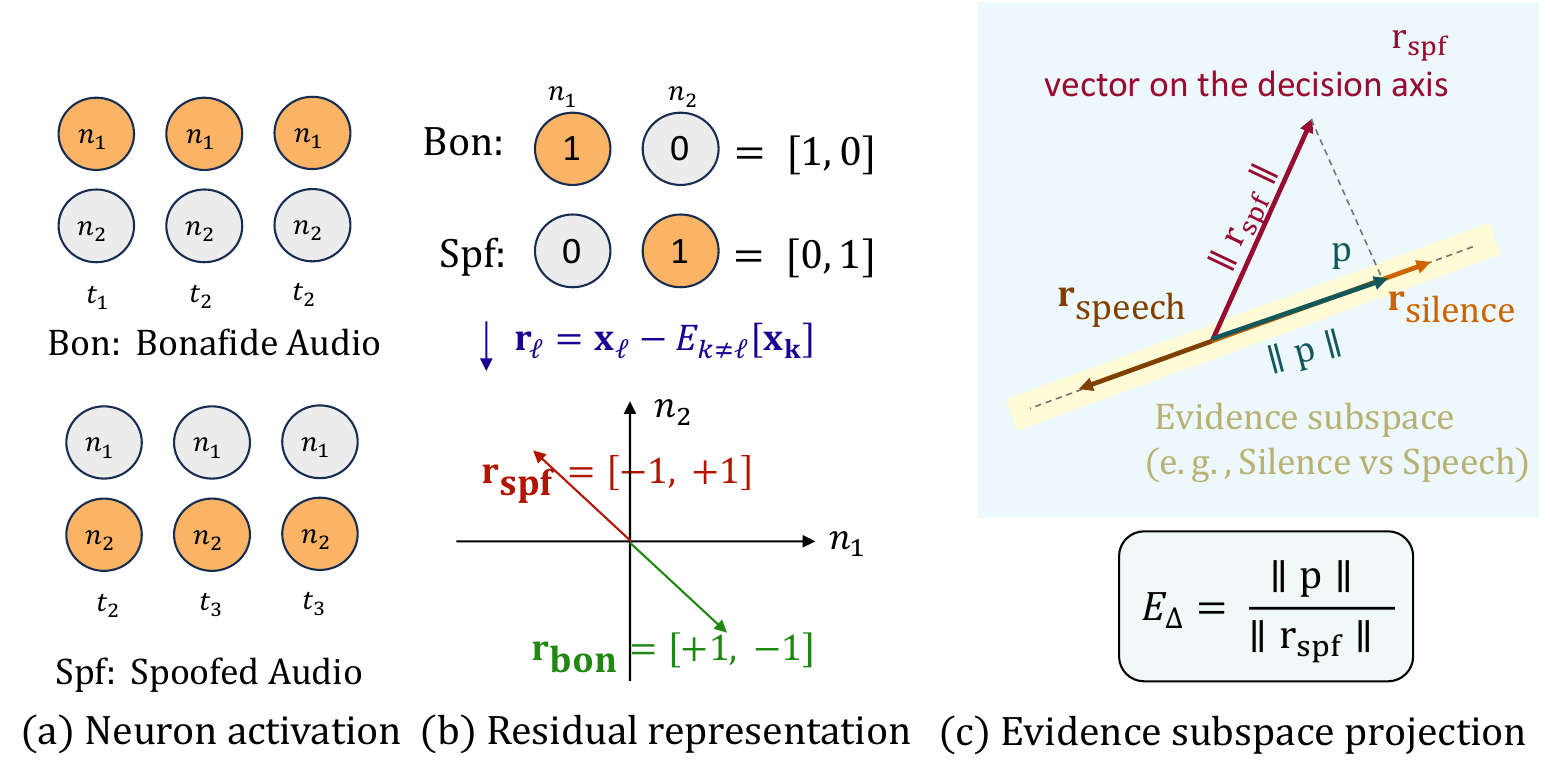}
    \caption{(a) Feed audio to SSL models to compute neuron activation probabilities. $n$ is a neuron and $t$ is a token. (b) Each label produces a neuron activation pattern. Subtracting the cross-label expected values yields one-vs-rest contrast vectors; in the binary case (bonafide/spoof), the two residuals are anti-parallel, and $\mathbf{r}_\text{spf}$ directly serves as the decision axis. (c) Projecting $\mathbf{r}_\text{spf}$ onto an evidence subspace (e.g., silence vs speech) gives $E_\Delta$: the fraction of the detection decision explained by that evidence factor.}
    \label{fig:overview}
\end{figure}
\section{Methods}

Our framework analyzes the SSL model's internal behavior by studying how feed-forward neurons respond to different groups. As shown in Fig.~\ref{fig:overview},
it first computes neuron activation statistics, then constructs residual representations in a shared neuron-activation space, and finally quantifies the explanatory power of each evidence group for the detection decision.

\subsection{Neuron Activation Probability}

SSL models such as XLSR\cite{babu22_xlsr2b} are based on Transformer~\cite{vaswani2017_attention, el2025comprehensive}.
Geva~\textit{et~al.}~\cite{geva2021transformer} propose viewing each Transformer FFN layer from a key-value perspective: 
in $\mathrm{FFN}(x) = f(xW_1^{\top})\,W_2$, the first weight matrix $W_1$ acts as keys and $W_2$ as values.
The output $f(xW_1^{\top}) \in \mathbb{R}^m$ then indicates how strongly each neuron is activated by the input. 
This perspective has been used to identify knowledge neurons~\cite{dai2022_knowledgeneurons}, language neurons~\cite{tang2024_languagespecificneurons}, and skill neurons~\cite{wang2022_skillneurons} in text-based models, and property neurons tied to phoneme categories and speaker attributes in speech SSL models~\cite{lin2024property}.

Inspired by their works, we first use a HuBERT quantizer to assign a numerical token label to each frame. These tokens encode fine-grained sub-phonetic information.
Then for one dataset $D$, let $\mathcal{G}$ denote a group of labels that share a common attribute (e.g., $\mathcal{G}_{\text{vocoder}} =
\{\text{MelGAN}, \text{HiFiGAN}, \ldots\}$), $g\in\mathcal{G}$ be one of the labels, $t$ a numerical token for a given frame, and $n$ a neuron in layer $l$, the conditional activation probability is

\begin{equation}
    P_g(n \mid t) =\frac{C_g(t,\, n)}{N_g(t)},
    \label{eq:cond_prob}
\end{equation}
where $C_g(t, n)$ is the co-occurrence count of token $t$ and neuron $n$ for label $g$, and $N_g(t) = \sum_{n'} C_g(t, n')$ is the total count of token $t$ for label $g$.
A co-occurrence is counted when neuron $n$'s activation value ranks in the top~1\% across all neurons for one frame tagged with token $t$.
$C_g(t,\, n)$ counts how many times neuron $n$ is activated for frames tagged with token $t$ under label $g$. Then $P_g(n \mid t)$ measures how consistently $n$ is activated. Under random activation, every neuron has a 1\% chance of being activated for any given frame, so the expected random baseline is  $\theta_{\mathrm{rand}} = 0.01$. A neuron is considered \emph{activated} for token $t$ belonging to label $g$ only if $P_g(n \mid t) > \theta_{\mathrm{rand}}$, i.e., it is activated more consistently than chance.
Then we define an indicator matrix $\mathbf{I} \in \{0, 1\}^{L\times|\mathcal{G}|\times|\mathcal{T}|\times N}$:

\begin{equation}
   \mathbf{I}_{l, g, t, n} = \mathbf{1} \left( P_g(n \mid t) > \theta_{\text{rand}} \right)
\end{equation}

To summarize how broadly a neuron responds across all tokens for label $g$, we finally define the \emph{activation coverage matrix}  $\mathbf{A} \in \mathbb{R}^{L\times|\mathcal{G}|\times N}$:

\begin{equation}
    \mathbf{A}_{l, g, n}=\frac{1}{|\mathcal{T}_g|} \sum_{t \in \mathcal{T}_g} \mathbf{I}[l, g, t, n]
    \label{eq:coverage}    
\end{equation}
where $\mathcal{T}_g$ is the set of distinct tokens for label $g$. A value close to 1 means the neuron responds to nearly all tokens for label $g$; a value near $\theta_{\text{rand}}$ indicates no label-specific preferences.

\subsection{Residual Representation}

We use $\mathbf{A}$ to construct a representation for each label in a shared neuron-activation space. For each dataset $D$ and label $g$, we concatenate activation-coverage values across all layers into $\mathbf{x}_{D,g} = [\mathbf{A}[1,g,:]; \ldots; \mathbf{A}[L,g,:]] \in \mathbb{R}^{d}$, where $d = L \cdot N$.
We assume $\mathbf{x}_{D,g} = \mathbf{c} + \mathbf{s}_{D} + \mathbf{e}_{g}$, where $\mathbf{c}$ is a component common to all conditions, $\mathbf{s}_{D}$ is a dataset-specific shift, and $\mathbf{e}_{g}$ is the label's effect. The \emph{residual representation} is obtained by subtracting the expected value over all other labels within the same dataset and group:

\begin{equation}
    \mathbf{r}_{g} = \mathbf{x}_{D,g} - \mathbb{E}_{g' \neq g}\bigl[\mathbf{x}_{D,g'}\bigr] = \mathbf{e}_{g} - \mathbb{E}_{g' \neq g}[\mathbf{e}_{g'}],
\end{equation}

Thus $r_g$ can be viewed as a \emph{one-vs-rest} contrast vector: the direction in the neuron-activation space that separates $g$ from the others.
In the binary case (e.g., bonafide vs spoof), the two residuals are exactly anti-parallel and lie on the decision axis (e.g., $r_{bon}-r_{spf}$). For multi-label groups (e.g., vocoder with $m$ labels), each $r_g$ captures the one-vs-rest decision direction for that label.
We then apply column-wise \emph{z-score} normalization and row-wise $L_2$-normalization to prevent any single neuron from dominating due to scale differences.

These representations operate in the \emph{neuron-activation space}, where each vector reflects how responsive the model's neurons are to a label.
This differs from feature-based methods that operate in the \emph{feature space}, where each point corresponds to a data sample. 
Moving in the neuron-activation space reflects a change in decision-making behavior, allowing us to study the interaction between model behavior and data distribution, rather than properties of the data alone.

\subsection{Evidence Subspace projection}

We define the \emph{decision group} $\mathcal{G_\text{decision}}=\{bonafide, spoof\}$, whose labels correspond to the classification target. We then define \emph{evidence groups}, each capturing an attribute whose relationship to the detection decision we aim to investigate. 
Since each residual representation is a one-vs-rest contrast vector, $\mathbf{r}_{\text{spf}}$ or $\mathbf{r}_{\text{bon}}$ directly captures the direction separating spoof and bonafide in the neuron-activation space, so either residual defines the decision axis. We use the normalized $\hat{\mathbf{r}}_{\text{spf}} = \mathbf{r}_{\text{spf}} / \lVert \mathbf{r}_{\text{spf}} \rVert$ as the decision axis. 
For a chosen evidence group with $m$ labels, the residual representations $\mathbf{F} = [\mathbf{r}_{\ell_1}, \mathbf{r}_{\ell_2},\ldots, \mathbf{r}_{\ell_m}]^\top \in \mathbb{R}^{m \times d}$ span a subspace capturing all one-vs-rest contrast directions within that group.
We obtain an orthonormal basis $\mathbf{Q} \in \mathbb{R}^{d \times r}$  via SVD ($\mathbf{F}^\top = \mathbf{U}\boldsymbol{\Sigma}\mathbf{V}^\top$), keeping the first $r$ columns of $\mathbf{U}$ corresponding to non-zero singular values. This rank selection guards against degenerate cases where evidence labels are highly correlated.
Projecting the decision axis onto this subspace yields the explanatory power:
\begin{equation}
     E_\Delta = \lVert\mathbf{Q}^\top \hat{\mathbf{r}}_{\text{spf}}\rVert^2
    \label{eq:edelta}
\end{equation}

$E_\Delta$ ranges from 0 to 1. A value close to 1 means the decision axis lies almost entirely within the evidence space, i.e., the model's one-vs-rest contrast for detecting spoof largely overlaps with its one-vs-rest contrasts among labels of that evidence group.
For fair cross-group comparison, we further normalized by the \emph{effective rank}\cite{roy2007_effectiverank} $\mathrm{erank}(\mathbf{F})$, since higher-rank subspace capture more energy by construction.
The resulting $E_\Delta/\text{erank}$ gives a per-effective-rank measure of the explanatory power.

\begin{figure*}[t]
    \centering
    \includegraphics[width=\textwidth]{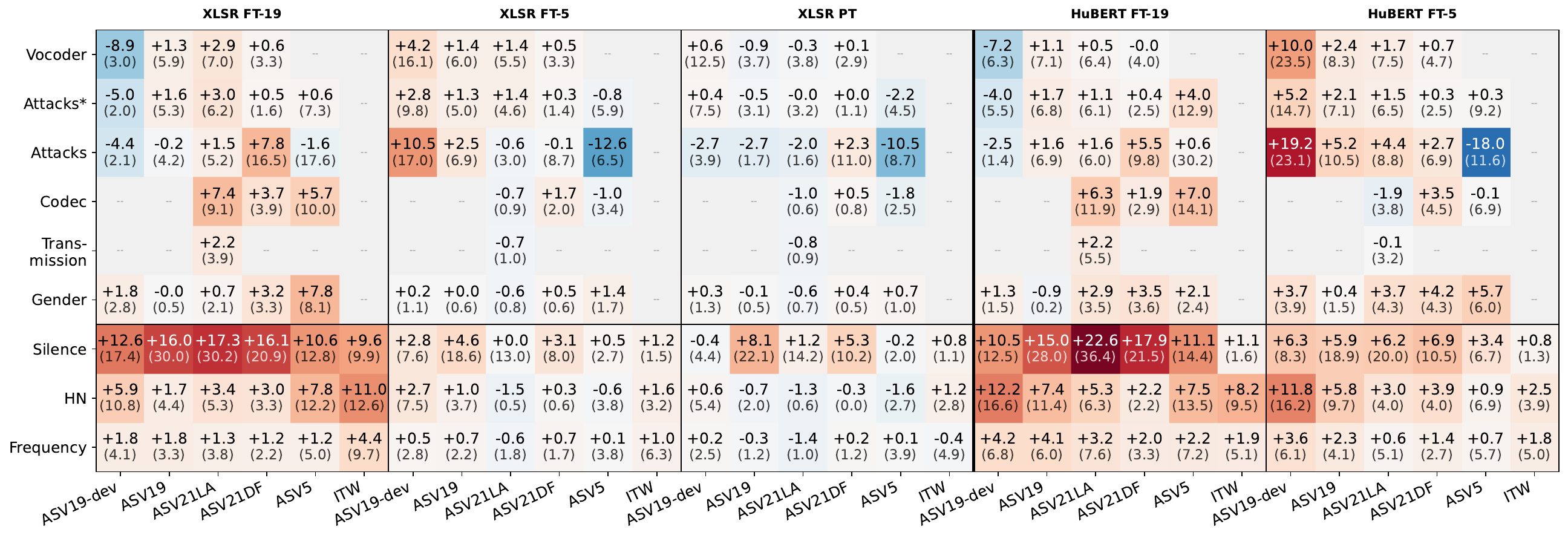}
    \caption{$E_\Delta/\text{erank}$ change relative to frozen models with absolute values.}
    \label{fig:fig3_mix_heatmap}
\end{figure*}

\section{Experiments}

We study 300M-parameter XLSR\cite{babu22_xlsr2b}
and HuBERT\cite{hsu2021_hubert}, each in three variants: frozen, fine-tuned on ASV19\cite{wang2020_asvspoof19} (denoted as FT-19), and fine-tuned on ASV5\cite{wang24_asvspoof5} (FT-5). We also include an XLSR model post-trained  (PT) on large-scale deepfake data\cite{ge2025_posttraining}; no comparable HuBERT PT model exists. For fine-tuning, we use a simple MLP backend\cite{el2025comprehensive} that projects SSL features to 128 dimensions, applies mean pooling, and maps to 2 classes; inputs are loop-padded to 4s  with no augmentation. Numerical tokens are obtained using HuBERT's quantizer 
\emph{L9 km500}.
We evaluate on six datasets including ASV19, ASV21LA\cite{yamagishi21_asvspoof21}, ASV21DF\cite{yamagishi21_asvspoof21}, and ASV5 test sets; since some overlap with XLSR PT's training data, we additionally include ASV19-dev and ITW\cite{muller22_generalize} as unseen sets.

We define nine evidence groups classified into two categories. 
(1)~\emph{Metadata-derived}: \textbf{Vocoder} (vocoder identity), \textbf{Attacks*} (dataset-specific attacker ID, e.g., A01, A02 in ASV19), \textbf{Attacks} (coarse attacker type: e.g., TTS for text-to-speech, VC for voice conversion), \textbf{Codec}, \textbf{Transmission} condition, and \textbf{Gender}, where available from ASVSpoof's metadata. (2)~\emph{Signal-level}: three frame-level labelings targeting cues identified in prior works.              
\textbf{Silence}: Leading/trailing silence has been shown to act as a shortcut; trimming it can increase EER from 3.6\% to 15.5\%~\cite{muller21_speechissilver}. 
 We label frames as ``silence'' if they belong to leading/trailing segments 30\,dB quieter than the maximum RMS; remaining frames are labeled as ``speech.''
\textbf{Frequency}: Subband analyses report mixed evidence on which bands are most discriminative~\cite{li2023_lowfreq, zhang2021_silencefusion, muttathu_2022voiced}. 
We compute the log-magnitude spectrogram, apply spectral whitening, and pool energy into four bands (0--2, 2--4, 4--6, 6--8\,kHz) via max-pooling. Each frame receives a multi-hot label if its pooled energy exceeds the frame-level mean by 0.5 standard deviations, yielding $2^4{=}16$ frequency-activation patterns.
\textbf{HN} (harmonic vs.\ noise): Deepfake detectors have been found to attend to specific noise-like (e.g., fricatives, stops)  and harmonic-dominant (e.g., nasals, vowels) phoneme classes~\cite{dhamyal2021_phonemicfeatures}. 
We use spectral flatness as a proxy: for non-silence frames, Otsu's threshold~\cite{otsu1979_threshold} separates \textit{harmonic-dominant} (low flatness) from \textit{noise-dominant} (high flatness) frames.

\section{Results and Discussion}

\subsection{What internal knowledge do frozen SSL models encode}

\begin{figure}[b]  
    \centering
    \includegraphics[width=\columnwidth]{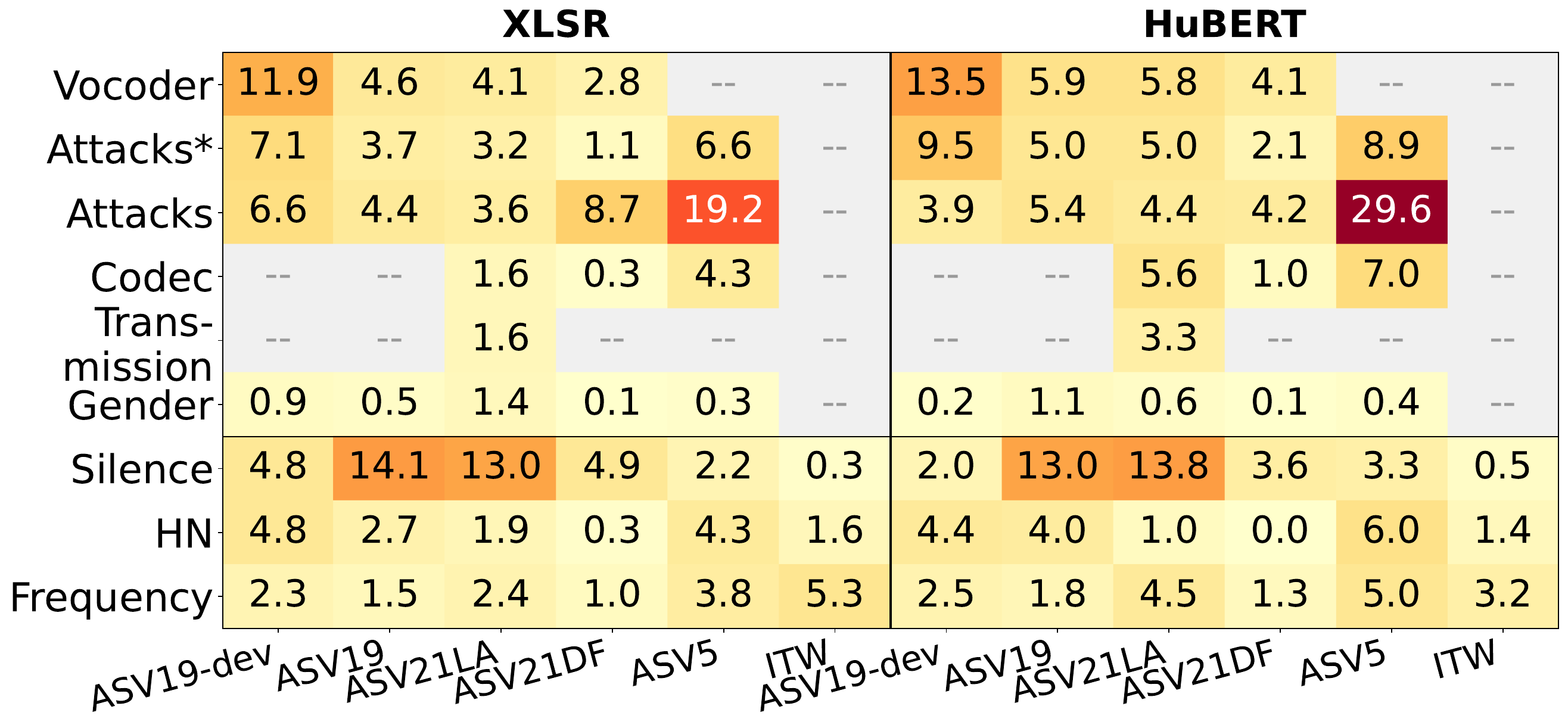}
    \caption{$E_\Delta/\text{erank}$ (\%) for frozen XLSR and HuBERT.}
    \label{fig:rq1}
\end{figure}

Fig.~\ref{fig:rq1} shows $E_\Delta/\text{erank}$ for all evidence groups.
Despite different pre-training strategies, XLSR and HuBERT produce similar heatmaps: relative rankings across evidence groups are largely preserved, indicating that the observed patterns reflect shared properties of SSL representations rather than model-specific artifacts.

\noindent\textbf{Signal-level category.}
On ASV19-test, the Silence group exceeds 10\% for both models, but drops below 0.5\% on ITW, confirming that silence cues are dataset-specific rather than general spoofing indicators~\cite{muller21_speechissilver}.
For Frequency and HN, no consistent cross-dataset pattern emerges: some group--dataset combinations reach 4--6\%, while others are near zero, suggesting these are also dataset-specific in frozen models.

\noindent\textbf{Metadata-derived category.}
Gender is nearly orthogonal to the decision axis, with $E_\Delta/\text{erank}$ below 1.4\% across all datasets and models, indicating that the neuron activation patterns distinguishing gender have little overlap with those used for deepfake detection.
For codec and transmission, which are more closely related to signal-level properties, HuBERT shows consistently higher $E_\Delta/\text{erank}$ than XLSR across all available datasets.

\noindent\textbf{Within-spoof groups.}
The Vocoder, Attacks*, and Attacks groups capture within-spoof variation computed on spoofed audio only. 
Each residual $\mathbf{r}_g$ within these groups is a one-vs-rest contrast vector among spoofed data.
Ideally, the decision axis should be orthogonal to these subspaces: if highly aligned, it means some spoofed data appear more similar to bonafide. For example, 
high $E_\Delta/\text{erank}$ for Attack group implies that some attack types appear more similar to bonafide than others; 
such a strategy is fragile and tends to misclassify some attacks as bonafide or bonafide as spoof.

For these groups, two cases stand out with unusually high values: 
Vocoder on ASV19-dev and Attacks on ASV5.
To analyze this, we conducted cos-similarity-based experiments.
For \textit{Vocoder on ASV19-dev}, the highest cosine similarity is between the \textit{Spectral Filtering+OLA} vocoder and $\mathbf{r}_\text{bon}$, 
reaching 0.608, which is significantly higher than the vocoder--bonafide average ($0.067 \pm 0.025$). The vocoder is used in A06, a source-filter voice conversion that reuses original residual signals \cite{wang2020_asvspoof19}.
We believe the reuse of original residual signals likely makes this vocoder a proxy for bonafide: the neuron activation patterns that distinguish it from other vocoders can be repurposed to distinguish bonafide from spoof.

For \textit{Attacks on ASV5}, a key difference is that a new group label \emph{adversarial attack} (AT) is introduced, referring to TTS combined with adversarial perturbations. The intuition is that detectors rely on attack-specific artifacts, so adversarial signals are added to disrupt this dependency and cause misclassification. As a result, we hypothesize AT might be ``aware'' of some TTS-specific artifacts and trigger neuron activation patterns similar to TTS.
To analyze, we compared cosine similarity between $\mathbf{r}_\text{attack}$ and $\mathbf{r}_{\text{bon}}$, and between $\mathbf{r}_\text{attack}$ and $\mathbf{r}_{\text{spf}}$. We found that TTS shows high similarity with $\mathbf{r}_\text{bon}$ (0.73 XLSR, 0.60 HuBERT), 
far above means ($0.15 \pm 0.36$ XLSR, $0.07 \pm 0.42$ HuBERT);
 AT points in the opposite direction, 
aligning with $\mathbf{r}_\text{spf}$ (0.56 XLSR, 0.70 HuBERT). In short, the detection decision axis aligns closer to TTS--AT detection axis on ASV5, causing higher $E_\Delta/\text{erank}$.

One explanation is that ASV5's advanced TTS produces speech closer to bonafide, while AT adds adversarial perturbations correlated \textit{strongly} with TTS artifacts. 
When computing the $\mathbf{r}_{\text{TTS}}$, shared artifact-related activation between TTS and AT is subtracted, leaving a representation closer to bonafide; for AT, less is removed, leaving $\mathbf{r}_{\text{AT}}$ aligned with spoof.
To verify, we construct modified residuals $\mathbf{r}'_\text{TTS}$ and $\mathbf{r}'_\text{AT}$ subtracting only $\mathbf{r}_\textbf{VC}$ to retain shared TTS artifacts. 
The TTS--bonafide similarities are 0.41 HuBERT and 0.28 XLSR, lower than $\mathbf{r}_\text{TTS}$ but still clearly higher than many others, confirming that ASV5 TTS appears closer to bonafide even without subtracting AT.
This alignment is stronger in HuBERT ($E_\Delta/\text{erank}$: 29.6\% vs.\ 19.1\% for XLSR), 
suggesting HuBERT may require more effort during training to correct this undesirable bias.

\subsection{What is the impact of different training strategies}

We fine-tune on ASV19 (low diversity, older methods) and ASV5 (high diversity, modern methods), representing two extremes. 
We include only an XLSR model post-trained (PT) on large-scale deepfake data since no comparable HuBERT post-trained model is available.
Fig.~\ref{fig:fig3_mix_heatmap} shows the change in $E_\Delta/\text{erank}$ relative to frozen models, alongside absolute values. Table~\ref{tab:rq2_train_test} reports EER.

\begin{table}[h]
\centering
\footnotesize
\newcommand{\ca}{\cellcolor{blue!8}}
\newcommand{\cb}{\cellcolor{orange!10}}
\caption{EER (\%) of model trained on \colorbox{blue!8}{\strut ASV19} or \colorbox{orange!10}{\strut ASV5}}
\label{tab:rq2_train_test}
\setlength{\tabcolsep}{3pt}
\begin{tabular}{@{}lcc@{\hspace{8pt}}lcc@{}}
\toprule
\ca Test & \ca XLSR & \ca HuBERT & \cb Test & \cb XLSR & \cb HuBERT \\
\midrule
\ca ASV19 & \ca 0.25 & \ca 0.53 & \cb ASV19   & \cb 15.75 & \cb 16.69 \\
\ca ASV21LA & \ca 5.37 & \ca 7.66 & \cb ASV21LA & \cb 19.33 & \cb 20.16 \\
\ca ASV21DF & \ca 3.81 & \ca 5.91 & \cb ASV21DF & \cb 10.86 & \cb 16.10 \\
\ca ASV5    & \ca 17.98 & \ca 22.47 & \cb ASV5  & \cb 5.69  & \cb 10.05 \\
\bottomrule
\end{tabular}
\end{table}

\begin{figure}[h]
    \centering
    \includegraphics[width=0.8\columnwidth]{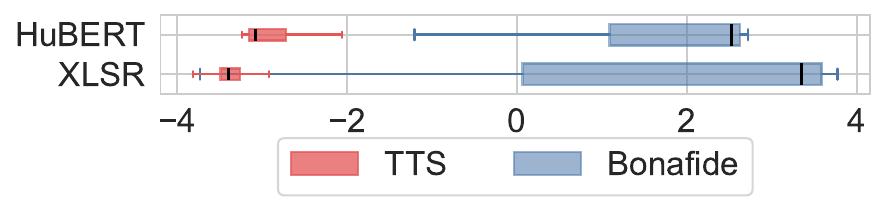}
    \caption{Bonafide score distributions on ASV21DF.}
    \label{fig:boxplot_tts_class1}
\end{figure}

\noindent\textbf{Within-spoof groups.}
Fine-tuning on either dataset reduces $E_\Delta/\text{erank}$ for Vocoder, Attacks*, and Attacks, indicating that training disentangles within-spoof variation from the detection axis. The reduction is especially large for HuBERT on ASV5 (up to 18\%), correcting the stronger prior alignment in frozen models. PT further reduces these values to even below frozen-model levels in most cases (e.g., Attacks on ASV19: frozen 4.4\%, PT 1.7\%).
XLSR achieves lower EER than HuBERT in all comparisons. We attribute HuBERT's weaker performance to its frozen model's stronger alignment with within-spoof variation, making it more difficult to ``unlearn'' the bias. Furthermore, we compare fine-tuned XLSR and HuBERT on each (training set, test set, within-spoof group) triplet: for each triplet, the model with lower $E_\Delta/\text{erank}$ is expected to achieve lower EER. This holds in 18 of 22 triplets. Three exceptions have $E_\Delta/\text{erank}$ differences below 2\%. One clear exception is FT-19 on ASV21DF for Attacks: XLSR has 6.8\% higher $E_\Delta/\text{erank}$ than HuBERT yet achieves lower EER (3.81\% vs.\ 5.91\%).
Analysis reveals this is a composition effect: VC dominates ASV21DF ($>$80\% of spoof samples) with low EER ($\sim$5\%), while TTS has $\sim$10\% EER.
The aggregated EER is pulled down by the well-classified VC majority.
Cosine similarity confirms: in XLSR FT-19, $\mathbf{r}_\text{TTS}$ has higher similarity with $\mathbf{r}_\text{bon}$ (0.4459) than in HuBERT (0.2433), indicating greater TTS--bonafide overlap, consistent with the score distributions in Fig.~\ref{fig:boxplot_tts_class1}. Hence although XLSR FT-19 has lower overall EER, it is weaker at detecting TTS compared to HuBERT FT-19.

\noindent\textbf{Gender.}~Finetuning increases gender $E_\Delta/\text{erank}$ on most datasets; PT restores near-orthogonality. However, the effect is different for finetuning: XLSR FT-19 (male:female=31:69) mostly produces larger increases than XLSR FT-5 (male:female=47:53), while for HuBERT the pattern reverses. This rules out marginal gender imbalance as the primary explanation. Since both SSL models are finetuned on identical data, the interaction suggests that gender might be entangled with other acoustic properties differently in each SSL, and finetuning surfaces these differences. 

\noindent\textbf{Signal-level variation.}
For Codec, Transmission, Silence, HN, and Frequency, FT-19 produces larger $E_\Delta/\text{erank}$ increases than FT-5, with Silence showing the largest gains. This reflects a fundamental difference in training data characteristics: ASV19's homogeneous recording conditions, silence structure, and spectral profiles allow signal-level properties to become confounded. ASV5's diversity decorrelates these properties from the detection task, preventing any single signal-level factor from dominating the decision axis.
The impact is different among models: HuBERT absorbs HN and Frequency into its decision axis more readily than XLSR under FT-5 (e.g., HN on ASV19-dev: HuBERT $+$11.8\% vs.\ XLSR $+$2.7\%), suggesting an SSL-dependent difference in sensitivity to spectral structure. PT suppresses HN and Frequency $E_\Delta/\text{erank}$ to near-frozen levels (all changes within $\pm$1.6\%) but only partially suppresses silence (ASV19: $+$8.1\%, ASV21DF: $+$5.3\%), indicating that the silence shortcut is persistent even under diverse training.

\section{Conclusion}

We propose evidence subspace projection to quantify which factors SSL front-ends rely on for deepfake detection. Frozen models already capture dataset-specific shortcuts such as silence and exhibit undesirable alignment between within-spoof variation and the detection axis. Fine-tuning reduces within-spoof alignment but amplifies signal-level shortcuts, especially under homogeneous training data (ASV19). Training data diversity (ASV5) decorrelates signal-level properties from the decision axis. Post-training further reduces within-spoof alignment and suppresses most signal-level dependencies, though silence remains partially persistent. Some explanatory power attributed to one group may originate from a correlated factor; future work should incorporate additional groups to further disentangle confounded effects.

\section{Generative AI Use Disclosure}
During the preparation of this work, the authors used ChatGPT to edit and polish the paper for improving readability, grammar, and phrasing. After using this tool, the authors reviewed and edited the content as needed and take full responsibility for the final content of the publication.

\bibliographystyle{IEEEtran}
\bibliography{mybib}

\begin{thebibliography}{10}
\providecommand{\url}[1]{#1}
\csname url@samestyle\endcsname
\providecommand{\newblock}{\relax}
\providecommand{\bibinfo}[2]{#2}
\providecommand{\BIBentrySTDinterwordspacing}{\spaceskip=0pt\relax}
\providecommand{\BIBentryALTinterwordstretchfactor}{4}
\providecommand{\BIBentryALTinterwordspacing}{\spaceskip=\fontdimen2\font plus
\BIBentryALTinterwordstretchfactor\fontdimen3\font minus
  \fontdimen4\font\relax}
\providecommand{\BIBforeignlanguage}[2]{{%
\expandafter\ifx\csname l@#1\endcsname\relax
\typeout{** WARNING: IEEEtran.bst: No hyphenation pattern has been}%
\typeout{** loaded for the language `#1'. Using the pattern for}%
\typeout{** the default language instead.}%
\else
\language=\csname l@#1\endcsname
\fi
#2}}
\providecommand{\BIBdecl}{\relax}
\BIBdecl

\bibitem{babu22_xlsr2b}
A.~Babu, C.~Wang, A.~Tjandra, K.~Lakhotia, Q.~Xu, N.~Goyal, K.~Singh, P.~{von
  Platen}, Y.~Saraf, J.~Pino, A.~Baevski, A.~Conneau, and M.~Auli, ``{XLS-R:
  Self-supervised Cross-lingual Speech Representation Learning at Scale},'' in
  \emph{{Proc. Interspeech}}, 2022, pp. 2278--2282.

\bibitem{pascu24_generalisable}
O.~Pascu, A.~Stan, D.~Oneata, E.~Oneata, and H.~Cucu, ``{Towards generalisable
  and calibrated audio deepfake detection with self-supervised
  representations},'' in \emph{{Proc. Interspeech}}, 2024, pp. 4828--4832.

\bibitem{combei25_AI4T}
D.~Combei, A.~Stan, D.~Oneata, N.~Müller, and H.~Cucu, ``{Unmasking real-world
  audio deepfakes: A data-centric approach},'' in \emph{{Proc. Interspeech}},
  2025, pp. 5343--5347.

\bibitem{ge2022_explaining}
W.~Ge, J.~Patino, M.~Todisco, and N.~Evans, ``{Explaining deep learning models
  for spoofing and deepfake detection with SHapley Additive exPlanations},'' in
  \emph{{IEEE International Conference on Acoustics, Speech and Signal
  Processing (ICASSP)}}.\hskip 1em plus 0.5em minus 0.4em\relax IEEE, 2022, pp.
  6387--6391.

\bibitem{liu24m_neuralspoofing}
T.~Liu, L.~Zhang, R.~K. Das, Y.~Ma, R.~Tao, and H.~Li, ``{How Do Neural
  Spoofing Countermeasures Detect Partially Spoofed Audio?}'' in \emph{{Proc.
  Interspeech}}, 2024, pp. 1105--1109.

\bibitem{muller21_speechissilver}
N.~Müller, F.~Dieckmann, P.~Czempin, R.~Canals, K.~Böttinger, and
  J.~Williams, ``{Speech is Silver, Silence is Golden: What do ASVspoof-trained
  Models Really Learn?}'' in \emph{{2021 Edition of the Automatic Speaker
  Verification and Spoofing Countermeasures Challenge}}, 2021, pp. 55--60.

\bibitem{shih2024_artifacts}
T.-H. Shih, C.-Y. Yeh, and M.-S. Chen, ``{Does Audio Deepfake Detection Rely on
  Artifacts?}'' in \emph{{IEEE International Conference on Acoustics, Speech
  and Signal Processing (ICASSP)}}, 2024, pp. 12\,446--12\,450.

\bibitem{stan25_interspeech}
A.~Stan, D.~Combei, D.~Oneata, and H.~Cucu, ``{TADA: Training-free Attribution
  and Out-of-Domain Detection of Audio Deepfakes},'' in \emph{{Proc.
  Interspeech}}, 2025, pp. 1543--1547.

\bibitem{geva2021transformer}
M.~Geva, R.~Schuster, J.~Berant, and O.~Levy, ``Transformer feed-forward layers
  are key-value memories,'' in \emph{Proceedings of the 2021 Conference on
  Empirical Methods in Natural Language Processing}, 2021, pp. 5484--5495.

\bibitem{lin2024property}
T.-Q. Lin, G.-T. Lin, H.-y. Lee, and H.~Tang, ``Property neurons in
  self-supervised speech transformers,'' in \emph{2024 IEEE Spoken Language
  Technology Workshop (SLT)}.\hskip 1em plus 0.5em minus 0.4em\relax IEEE,
  2024, pp. 401--408.

\bibitem{vaswani2017_attention}
A.~Vaswani, N.~Shazeer, N.~Parmar, J.~Uszkoreit, L.~Jones, A.~N. Gomez,
  {\L}.~Kaiser, and I.~Polosukhin, ``{Attention Is All You Need},'' in
  \emph{{Advances in Neural Information Processing Systems (NeurIPS)}}, 2017,
  pp. 6000--6010.

\bibitem{el2025comprehensive}
Y.~El~Kheir, Y.~Samih, S.~Maharjan, T.~Polzehl, and S.~M{\"o}ller,
  ``Comprehensive layer-wise analysis of ssl models for audio deepfake
  detection,'' in \emph{Findings of the Association for Computational
  Linguistics: NAACL 2025}, 2025, pp. 4070--4082.

\bibitem{dai2022_knowledgeneurons}
D.~Dai, L.~Dong, Y.~Hao, Z.~Sui, B.~Chang, and F.~Wei, ``{Knowledge Neurons in
  Pretrained Transformers},'' in \emph{{Proceedings of the 60th Annual Meeting
  of the Association for Computational Linguistics (ACL)}}, 2022, pp.
  8493--8502.

\bibitem{tang2024_languagespecificneurons}
T.~Tang, W.~Luo, H.~Huang, D.~Zhang, X.~Wang, X.~Zhao, F.~Wei, and J.-R. Wen,
  ``{Language-Specific Neurons: The Key to Multilingual Capabilities in Large
  Language Models},'' in \emph{{Proceedings of the 62nd Annual Meeting of the
  Association for Computational Linguistics (ACL)}}, 2024, pp. 5701--5715.

\bibitem{wang2022_skillneurons}
X.~Wang, K.~Wen, Z.~Zhang, L.~Hou, Z.~Liu, and J.~Li, ``{Finding Skill Neurons
  in Pre-trained Transformer-based Language Models},'' in \emph{{Proceedings of
  the 2022 Conference on Empirical Methods in Natural Language Processing
  (EMNLP)}}, 2022, pp. 11\,132--11\,152.

\bibitem{roy2007_effectiverank}
O.~Roy and M.~Vetterli, ``{The effective rank: A measure of effective
  dimensionality},'' in \emph{{European Signal Processing Conference
  (EUSIPCO)}}.\hskip 1em plus 0.5em minus 0.4em\relax IEEE, 2007, pp. 606--610.

\bibitem{hsu2021_hubert}
W.-N. Hsu, B.~Bolte, Y.-H.~H. Tsai, K.~Lakhotia, R.~Salakhutdinov, and
  A.~Mohamed, ``{HuBERT: Self-Supervised Speech Representation Learning by
  Masked Prediction of Hidden Units},'' \emph{{IEEE/ACM Transactions on Audio,
  Speech, and Language Processing}}, vol.~29, pp. 3451--3460, 2021.

\bibitem{wang2020_asvspoof19}
X.~Wang, J.~Yamagishi, M.~Todisco, H.~Delgado, A.~Nautsch, N.~Evans,
  M.~Sahidullah, V.~Vestman, T.~Kinnunen, K.~A. Lee \emph{et~al.}, ``{ASVspoof
  2019: A large-scale public database of synthesized, converted and replayed
  speech},'' \emph{{Computer Speech \& Language}}, vol.~64, p. 101114, 2020.

\bibitem{wang24_asvspoof5}
X.~Wang, H.~Delgado, H.~Tak, J.~weon Jung, H.~jin Shim, M.~Todisco, I.~Kukanov,
  X.~Liu, M.~Sahidullah, T.~H. Kinnunen, N.~Evans, K.~A. Lee, and J.~Yamagishi,
  ``{ASVspoof 5: crowdsourced speech data, deepfakes, and adversarial attacks
  at scale},'' in \emph{{The Automatic Speaker Verification Spoofing
  Countermeasures Workshop (ASVspoof 2024)}}, 2024, pp. 1--8.

\bibitem{ge2025_posttraining}
W.~Ge, X.~Wang, X.~Liu, and J.~Yamagishi, ``{Post-training for Deepfake Speech
  Detection},'' \emph{arXiv preprint arXiv:2506.21090}, 2025.

\bibitem{yamagishi21_asvspoof21}
J.~Yamagishi, X.~Wang, M.~Todisco, M.~Sahidullah, J.~Patino, A.~Nautsch,
  X.~Liu, K.~A. Lee, T.~Kinnunen, N.~Evans, and H.~Delgado, ``{ASVspoof 2021:
  accelerating progress in spoofed and deepfake speech detection},'' in
  \emph{{2021 Edition of the Automatic Speaker Verification and Spoofing
  Countermeasures Challenge}}, 2021, pp. 47--54.

\bibitem{muller22_generalize}
N.~Müller, P.~Czempin, F.~Diekmann, A.~Froghyar, and K.~Böttinger, ``{Does
  Audio Deepfake Detection Generalize?}'' in \emph{{Proc. Interspeech}}, 2022,
  pp. 2783--2787.

\bibitem{li2023_lowfreq}
M.~Li and X.-P. Zhang, ``{Robust Audio Anti-Spoofing System Based on
  Low-Frequency Sub-Band Information},'' in \emph{{IEEE Workshop on
  Applications of Signal Processing to Audio and Acoustics (WASPAA)}}, 2023,
  pp. 1--5.

\bibitem{zhang2021_silencefusion}
Y.~Zhang, W.~Wang, and P.~Zhang, ``{The Effect of Silence and Dual-Band Fusion
  in Anti-Spoofing System},'' in \emph{{Proc. Interspeech}}, 2021, pp.
  4279--4283.

\bibitem{muttathu_2022voiced}
A.~S. M.~S. Pillai, P.~L.~D. Leon, and U.~Roedig, ``{Detection of voice
  conversion spoofing attacks using voiced speech},'' in \emph{{Nordic
  Conference on Secure IT Systems (NordSec)}}, 2022, pp. 159--175.

\bibitem{dhamyal2021_phonemicfeatures}
H.~Dhamyal, A.~Ali, I.~A. Qazi, and A.~A. Raza, ``{Using Self Attention DNNs to
  Discover Phonemic Features for Audio Deep Fake Detection},'' in \emph{{IEEE
  Automatic Speech Recognition and Understanding Workshop (ASRU)}}, 2021, pp.
  1178--1184.

\bibitem{otsu1979_threshold}
N.~Otsu, ``{A Threshold Selection Method from Gray-Level Histograms},''
  \emph{{IEEE Transactions on Systems, Man, and Cybernetics}}, vol.~9, no.~1,
  pp. 62--66, 1979.

\end{thebibliography}

\end{document}